# Overview of Sensing Attacks on Autonomous Vehicle Technologies and Impact on Traffic Flow

Zihao Li
*Zachry Department of Civil & Environmental Engineering*
*Texas A&M University*
College Station, Texas, USA
scottlzh@tamu.edu

Sixu Li
*Department of Multidisciplinary Engineering*
*Texas A&M Univeristy*
College Station, Texas, USA
sixuli@tamu.edu

Hao Zhang
*Zachry Department of Civil & Environmental Engineering*
*Texas A&M University*
College Station, Texas, USA
zhanghao230@tamu.edu

Yang Zhou*, Ph.D.
*Zachry Department of Civil & Environmental Engineering*
*Texas A&M University*
College Station, Texas, USA
yangzhou295@tamu.edu

Siyang Xie, Ph.D.
*Cruise LLC*
San Francisco, California, USA
siyang.xie09@gmail.com

Yunlong Zhang, Ph.D.
*Zachry Department of Civil & Environmental Engineering*
*Texas A&M University*
College Station, Texas, USA
yzhang@civil.tamu.edu

*Abstract*— While perception systems in Connected and Autonomous Vehicles (CAVs), which encompass both communication technologies and advanced sensors, promise to significantly reduce human driving errors, they also expose CAVs to a variety of cyberattacks. These include both communication and sensing attacks, which have the potential to jeopardize not only individual vehicles but also overall traffic safety and efficiency. While much research has focused on communication attacks, sensing attacks, which are equally critical, have garnered less attention. To address this gap, this study offers a comprehensive review of potential sensing attacks and their impact on target vehicles, focusing on commonly deployed sensors in CAVs such as cameras, LiDAR, Radar, ultrasonic sensors, and GPS. Based on this review, we discuss the feasibility of integrating hardware-in-the-loop experiments with microscopic traffic simulations. We also design baseline scenarios to analyze the macro-level impact of sensing attacks on traffic flow. The aim of this study is to bridge the research gap between individual vehicle sensing attacks and broader macroscopic impacts, thereby laying the foundation for future systemic understanding and mitigation.

## I. INTRODUCTION

Connected and Autonomous Vehicles (CAVs) have been heralded as a transformative solution for addressing enduring challenges in transportation, including traffic safety, operational efficiency, and fuel consumption [1], [2]. Utilizing cutting-edge communication and sensor technologies, these vehicles perceive their surrounding environment, enabling onboard computing systems to process received information through decision-making algorithms. This results in autonomous vehicle navigation, including acceleration, braking, and steering, thus eliminating the need for human intervention [3]. A growing body of research [4] demonstrated that CAVs can substantially mitigate human driving errors by leveraging real-time sensor data, mitigating factors such as fatigue and distraction, and making prompt, rational decisions. Despite their advanced safety features and efficiencies, CAVs are not impervious to risks. The perception systems in CAV are particularly susceptible to being attacked due to their reliance on various sensors and communication technologies.

The perception systems in CAVs are crucial for their safe and efficient operation. Any compromise in these systems can have severe implications because subsequent decision-making algorithms rely on information gleaned from the perception systems, such as the spacing, velocity, and acceleration of preceding vehicles, to determine the vehicle's actions. The cyberattack in CAV's perception system can be categorized into two types, including communication attack and sensing attack [5]. Communication attacks refer to malicious activities that target the communication layers used for vehicle-to-everything (V2X) communication. The potential communication attacks have been categorized into three main types (i.e., falsification, delay/replay, and collusion attack) and the corresponding impact on safety and stability in vehicular platoon has been evaluated with the help of the extended Intelligent Driver Model [5]. Although this study exclusively investigated the longitudinal impact of cyberattacks in pure CAV platoons, additional studies have since been conducted to analyze the impact of communication attacks under various scenarios, for example, minor cyberattack [6], cyberattack considering lane changes [7], cyberattack-induced cooperative adaptive cruise control (CACC) degradation [8], and cyberattack under heterogeneous traffic flow [9]. The numerical experiment results from the aforementioned studies have demonstrated that cyberattacks on CAV may disturb traffic flow and cause vehicles unnecessary delays, abrupt acceleration/deceleration as well as rear-end collisions. The existing research on communication attacks has commonly assumed that the onboard sensors of CAV are reliable and robust, so only the information in communication layers is vulnerable to be attacked. However, multiple prior research has reported sensing attacks in CAV, such as camera blinding [10], adversarial attack for LiDAR [11] and Radar system [12]. The cyberattack in the sensing layers can also compromise the security and performance of autonomous driving functions and potentially lead to a safety risk or even a crash. Existing research mainly focuses on sensor-specific attack strategies for target CAV, lacking a comprehensive review of its corresponding impact on traffic flow. To address this gap, this study aims to provide an exhaustive overview of sensing attack models and their potential impact. The goal is to establish baseline scenarios for sensing attacks that can serve as a reference for testing countermeasures and for evaluating the macroscopic, system-level (i.e., traffic flow) impacts of these attacks beyond just the targeted vehicles. Moreover, the future direction for sensing attacks at transportation operations is discussed.

## II. POTENTIAL SENSING ATTACKS IN CONNECTED AND AUTONOMOUS VEHICLES

In this section, we provide an overview of potential sensing attacks targeting a range of sensors in CAV systems, which covers both the attack models and the consequences. While the types of sensors deployed in CAVs can vary by manufacturer and specific vehicle model, commonly used sensors include cameras, Light Detection and Ranging (LiDAR), Radio Detection and Ranging (RADAR), ultrasonic sensors, and Global Positioning Systems (GPS). These sensors function either independently or in collaboration to execute various automation tasks, such as car following, lane departure warning, and lane-changing maneuvers [13]. The mapping between sensors and their corresponding automation function is illustrated in Fig. 1.

### A. Camera

Within CAVs, cameras are the fundamental sensor, which serves multiple automation functions such as object detection, lane detection, and traffic sign reading, allowing the CAV to navigate safely and make informed maneuver decisions. Additionally, cameras work in conjunction with other sensors like LiDAR and Radar to provide a comprehensive view of the vehicle's surroundings and to determine the distance and relative speed of the preceding vehicle by comparing the picture in different timesteps. Traditionally, cameras are vulnerable to both *physical tampering*, such as lens obstruction, and *electronic interference*, such as exposure to intense light or laser beams that can cause image overexposure or temporary blinding. However, these two approaches are easily detected [14]. CAV always requires the imagine-processing algorithms (e.g., deep convolutional neural network) to analyze the collected image or video data, so the malicious attacker may manipulate the algorithm's input data to deceive its machine learning algorithm without causing any physical tampering or setting of transitional security alerts [15]. In the context of cameras, it could involve subtly altering the pixel in an image in a way that causes the CAV to misidentify the object, such as traffic control sign. Examples of such *physical adversarial attacks* include modifying traffic signs with stickers, visible light, infrared laser, or shadow projection [16], [17] causing the algorithms to misunderstand the surrounding environment. Instead of misidentifying the traffic control sign, Cao et al. [18] had designed a sensing attack to generate a physically realizable adversarial 3D-printed object that could mislead a multi-sensor fusion-based perception (i.e., camera and LiDAR).

However, Lu et al. [19] asserted that when the camera can view objects from different distances and at different angles, the adversarial objects do not disrupt object detection from a moving platform or just misidentify the object from a short distance. Later, Gao et al., [20] proposed an end-to-end adversarial patch generation for the preceding vehicle. Experiments were conducted using the CARLA driving simulator in conjunction with OpenPilot. The results indicate that adversarial attacks are viable against moving CAVs and could potentially lead to speed instability and even rear-end collisions.

### B. LiDAR

LiDAR sensors emit laser beams and measure the time it takes for the beams to bounce back after hitting an object. LiDAR data is then used to create a 3D point cloud, providing a detailed and accurate (range accuracy of 0.5 to 10mm) representation of the surrounding environment. Unlike cameras, LiDAR is less affected by adverse weather or lighting conditions, making it a reliable part of a CAV's sensor. The categories of sensing attacks on LiDAR could be divided into *jamming* and *spoofing* [21].

Under jamming attacks, LiDAR is sensitive to low-cost near-infrared or infrared light sources to generate the fake dots in point cloud data, but these light sources cannot blind the LiDAR. However, the low-budget low-power visible lasers can easily temporarily blind the LiDAR [22], for example, saturating and Denies of Service (DOS). The malicious attacker could record the signal sent by LiDAR and replay the signal at a later point in time (replay attack) or transmit the recorded signal to another CAV (relay attack) to mislead the sensor to generate the fake dots and further misidentify the non-existence object. Such an attack is relatively hard to achieve in the real world because of the restriction for receiving angle [21]. Gao et al., [11] first performed the LiDAR-based attack, and they found the attack performance is undesirable if blindly applied the spoofing attack, so they formulated the attack as an optimization problem to enhance the attack success rate. In the simulation, they fooled the victim CAV into making an emergency brake due to the spoofed obstacles. In Shin et al. [21], they induced several fake data points by relay attack to deceive the CAV into believing that an object was closer than it actually was. Considering the cooperative perception of the camera and LiDAR, Hallyburton et al. [23] had designed a context-based attack (frustum attack), which could simultaneously spoof both camera and LiDAR semantics and cause CAV to mistakenly determine the spacing and velocity of preceding vehicles.

### C. Radar

Radar sensors emit electromagnetic signals to gauge distance and velocity, similar to the function of LiDAR but with the added advantage of better performance in adverse weather conditions like rain, fog, or snow. While Radar offers robust low-light operation, it generally provides lower resolution compared to LiDAR. Most CAV radars employ frequency-modulated continuous wave (FMCW) technology within the millimeter wave (mmW) frequency band. These radars serve different functions depending on their range: long-range radars are used for car-following operations, mid-range radars aid in lane-changing and lane-keeping, and short-range radars collaborate with ultrasonic sensors for parking assistance. The primary types of attacks that target Radar systems are *Jamming* and *Spoofing* [24]. Jamming attacks intentionally transmit radio frequency signals in the radar's operating frequency band to overload or interfere with the receiver, thereby reducing the signal-to-noise ratio. Common jamming techniques utilize continuous-wave signals, random noise, or modulated signals to disrupt radar functionality [25]. On the other hand, spoofing attacks involve the retransmission of radar signals to feed false information to the radar system, thereby misleading its object detection capabilities. For a spoofing attack to be effective, the spoofing device must be strategically placed in the same lane as the targeted radar, owing to the direction-of-arrival detection capability of automotive radars. Such an attack could generate an artificial object within an unsafe distance, thereby triggering false alarms in the victim vehicle's collision-avoidance system. Spoofing attacks from a different lane are generally detectable due to the wider angle of arrival [26]. Komissarov and Wool [27] used a single rogue radar to physically-coherently spoof the distance and velocity measured by the victim vehicle, presenting phantom measurement and resulting in the emergency stop and abrupt acceleration. Yan et al. [28] successfully executed jamming and spoofing attacks on the radar system of a Tesla Model S led to the introduction of a "ghost vehicle" and caused disruptions in accurately measuring the distance from the preceding vehicle.

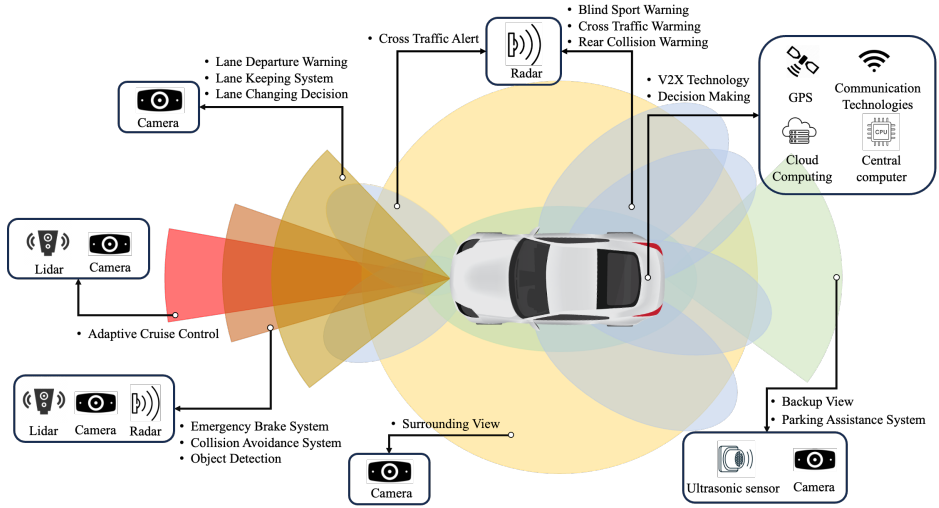

Fig. 1. Deployed Sensors and its Corresponding Automation Functions in CAV

*D. Ultrasonic sensors*

Ultrasonic sensors employ ultrasonic pulses to detect objects and measure their distance, especially for close-range detection. Typically, the ultrasound frequency falls within the range of 40 to 50kHz. These sensors feature a transducer responsible for both emitting and receiving ultrasonic waves. Attacks on ultrasonic sensors are typically categorized into two main groups: *physical signal level attacks* and *sensor hardware level attacks* [29]. Physical signal level attacks involve altering the sensor's environment to disrupt its operation. Sensor hardware level attacks, on the other hand, target the data collection or processing functions of the sensor. Lim et al. conducted a test by covering both the transmitter and receiver of the sensor and masking an obstacle with acoustic foam [30]. Most existing studies have focused on physical signal level attacks, encompassing two well-known methods: *jamming* and *spoofing*. Jamming employs noisy ultrasound signals to disrupt the normal operation of the sensor, while spoofing involves using meticulously designed ultrasounds to falsify the presence of obstacles. Numerous research efforts have been devoted to developing attack models for ultrasonic sensors and analyzing the impact of these methods [31], [32]. Given that ultrasonic sensors are primarily employed to detect obstacles in blind spots and measure distances to nearby objects, attacks on these sensors can have severe consequences. Such attacks might cause vehicles to continue moving when they should stop or halt when they should continue. Additionally, the compromised blind spot detection system could elevate the risk of collisions during lane-changing or merging maneuvers.

*E. GPS receiver*

GPS sensors operate by transmitting and receiving signals from satellites orbiting the Earth, enabling precise vehicle location determination. Attacks on GPS sensors primarily fall into two categories: *jamming* and *spoofing* [33]. Jamming involves the injection of noise signals into GPS receivers to disrupt the reception of authentic signals. Spoofing, on the other hand, entails the creation of meticulously crafted signals to deceive GPS receivers. Numerous research efforts have delved into the study of GPS sensor vulnerabilities and methods to bolster their defenses. However, a significant portion of studies has concentrated on GPS systems for unmanned aerial vehicles [34], [35], with only a limited focus on those for autonomous vehicles [36]. The investigations on autonomous vehicles typically involve the analysis of attack models against GPS and propose strategies to enhance their security. For instance, Sanders et al. developed a GPS attacker model for autonomous vehicles and proposed an effective framework for localizing attackers. This framework underwent testing under both static and mobile spoofer scenarios [37]. Yang et al. detected spoofing attacks on GPS by combining data-driven and model-driven methodologies [38]. Attacks on GPS systems in autonomous vehicles can have far-reaching consequences. They can compromise the accuracy of vehicle positioning and, in some cases, deliberately provide false location data. This disruption can disable crucial functions of autonomous vehicles, such as lane-keeping and route planning, which rely heavily on high-definition mapping and precise GPS positioning. Furthermore, these attacks pose significant safety risks, including vehicle collisions and crashes involving pedestrians near roadways.

## III. SIMULATION & EXPERIMENT

To fully overview the potential sensing attack in CAVs and its corresponding impact, the simulation and experiment designed should be examined, which is critical for the validation and evaluation of sensing attack. Experimental setups for sensing attacks in prevailing studies can be grouped into two categories: simulation and hardware-in-the-loop experiments (i.e., sensor and vehicle).

Simulation experiments are typically conducted using open-source simulation platforms and/or open-source autonomous driving software. Cao et al. [18] utilized the Baidu Apollo software in conjunction with the Unity-based LGSVL simulator for end-to-end safety simulations. The raw data for cameras and LiDAR are sourced from the simulation engine. In another study, Cao et al. [39] assessed the impact of a LiDAR spoofing attack on driving decisions using the simulation feature provided by Baidu Apollo called Sim-control, which formulates decisions from real-world sensor data. Kapoor et al. [40] injected radar spoofing attack signals into the car-following model in the ACC package of Matlab to evaluate short-term static radar attacks, long-term static radar attacks, and sinusoidal radar attacks. You et al. [41] harnessed a public nuScenes dataset, producing 3D point-clouds from 10 driving scenes. They utilized detection models PointPillars and SECOND, along with MotionNet for object-motion prediction, to scrutinize a LiDAR spoofing attack detection approach. While simulation experiments offer cost-effective and straightforward evaluations of full-stack autonomous driving software, they often don't emulate real-time data from actual sensors or authentically depict AV responses under attack conditions.

Sensor-in-the-loop experiments commonly employ offboard sensors to simulate sensing attacks against onboard AV sensors. Cao et al. [18] conducted a scaled-down experiment using Velodyne VLP-16 LiDAR and an iPhone 8 Plus rear camera in a miniaturized road setting to capture adversarial object detection results. In another study, Cao et al. [39] targeted the VLP-16 PUCK LiDAR System by Velodyne, injecting counterfeit data to assess the attack's effect on the sensing module. Xu et al. [29] examined the security of ultrasonic sensors for autonomous vehicles by launching random and adaptive spoofing attacks, as well as jamming attacks against standalone ultrasonic sensors. While these experiments can produce genuine sensing results, they typically involve stationary attackers and sensors. The dynamic challenge of targeting sensors on moving vehicles with precision is often overlooked, leading to a potential underrepresentation of real-world attack success rates.

Vehicle-in-the-loop experiments engage real vehicles to yield more authentic sensing outcomes. Cao et al. [18] utilized a vehicle outfitted with the high-end Velodyne HDL-64E LiDAR, conducting adversarial object detection tests on actual roads by manually driving the vehicle. Xu et al. [29] carried out random spoofing and jamming attacks against the integrated ultrasonic sensors of an autopilot-equipped Tesla Model S, and similarly targeted ultrasonic sensors in various vehicle models with driving assistance systems. Their aim was to probe the security of ultrasonic sensors in autonomous vehicles. Sun et al. [42] orchestrated software-in-the-loop experiments where a Lincoln MKZ, fitted with a TI IWR6843 radar, was targeted by a static attacker from the roadside. The compromised detection results were then fed into the Baidu Apollo software for decision-making. While vehicle-involved experiments offer insights from real-world traffic settings, safety precautions often limit them to studying only the sensing modules. As a result, the full scope of a sensing attack's impact on autonomous vehicle behavior in real-world conditions might not be completely captured.

Moreover, current research predominantly centers on the vulnerabilities and repercussions for individual vehicles, neglecting the potential impact on vehicular platoons or overall traffic systems. This gap is especially concerning for transportation engineers and policymakers who need to understand the systemic implications of sensing attacks. To bridge this gap, integrating microscopic traffic simulation platforms like SUMO, VISSIM, or AIMSUN with open-source autonomous driving software or hardware-in-the-loop experiments could be invaluable. These simulations can model how disturbances such as velocity instability or abrupt acceleration and deceleration ripple through upstream vehicles, thereby capturing the macroscopic impact. The importance of human factors in sensing attacks cannot be overstated, as human intervention can override CAV systems and alter driving behavior. Incorporating driving simulators into the testing framework could offer a more nuanced understanding of the impact of sensing attacks in mixed traffic conditions and enable a more accurate assessment of their consequences. In addition, experiments involving real-world sensing attacks should be conducted in controlled settings to assess an operating CAV's most realistic response to various types of sensing attacks under different scenarios (e.g., weather and light conditions) or maneuvers (e.g., car following and lane changing maneuver).

IV. CORRESPONDING IMPACT ON TRAFFIC FLOW

In this section, we designed baseline scenarios intended for future research. These scenarios will serve as the foundation for analyzing the impact of sensing attacks and for evaluating the efficacy of strategies aimed at mitigating these attacks. Different from a communication attack, the sensing attack has the following properties: i) physically consistent; ii) mainly target in spacing and velocity, instead of acceleration; iii) attack success rate is time-varying and related to attack magnitude.

Due to the direction-of-arrival detection capabilities in advanced sensors like Radar and LiDAR, executing a successful attack from a vehicle in a separate lane presents additional challenges [26]. As a result, the scenarios we design primarily focus on car-following situations. In general, sensing attacks can be categorized into two types: jamming/saturating and spoofing. Sensor jamming is generally easier to detect, as the sensor fails to receive any information. Upon detecting such an attack, the target vehicle issues an initial warning. This prompts a period of reaction time for the human driver to take control. During this time, the vehicle is likely to maintain its existing speed. After transitioning to human-driving mode for a specified duration, vehicle control can be safely returned to CAV, according to the human driver's judgment, as depicted in Fig. 2(a). The second type of attack is sensor spoofing, which is more difficult to detect and does not trigger a warning alert for the human driver. Unlike jamming, the objective of sensor spoofing is to deceive the target vehicle by creating a false environment. This exploits the semantic gap between the actual conditions and how the sensor interprets them. As a result, spoofing can manifest in two main ways: either by blinding the sensor or by creating a "ghost" vehicle. In Fig. 2(b), the spoofing attack misleads the target vehicle into believing that no preceding vehicle exists, when in fact one does. It's important to note that the success rate of the attack can vary based on real-time conditions, as the target vehicle is in motion [19]. In other words, the target vehicle may intermittently detect the presence of a preceding vehicle, experiencing it as appearing and disappearing. This instability could undermine human trust in autonomous vehicles, potentially leading to a downgrade from autonomous to human-driven operations. Another form of spoofing attack, depicted in Fig. 2(c), does not directly blind the target vehicle. Instead, it generates instability in the spacing or velocity of the preceding vehicle by creating a "ghost vehicle." This instability mainly manifests in two forms: constant attacks, which produce a stable offset from the true state, and random attacks, which introduce a random offset. Although false states introduced by the attack can potentially be detected or reconstructed using anomaly detection algorithms [43] and sensor fusion technologies [44], the rate of detection and quality of reconstruction are heavily influenced by the magnitude of the attack [11]. Typically, the detection rate and quality of reconstruction decline as the attack magnitude increases, but the precise relationship remains unknown.

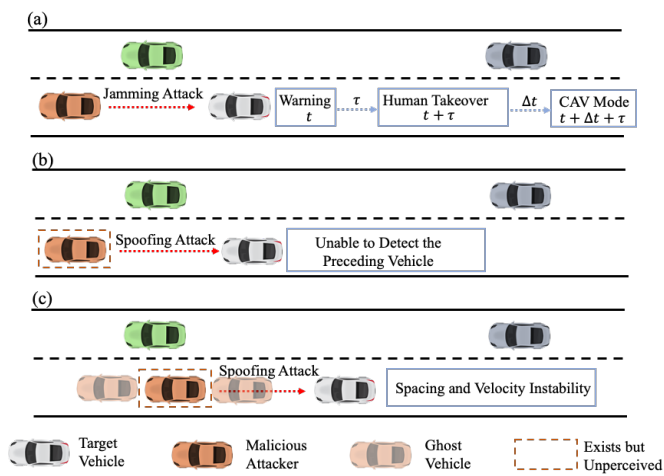

Fig. 2. Designed Sensing Attack Senarios

## V. FUTURE DIRECTION & CONCLUSION

In the preceding section, we outlined typical sensing attack scenarios and highlighted their potential impact on traffic flow. These designed scenarios can serve as a baseline for assessing the influence of sensing attacks on vehicular platoons and traffic flow, as well as for testing the performance of robust CAV controller in presence of cyberattacks. To enable a more comprehensive impact analysis, future research could incorporate the following factors inspired by the discussion above: i) the relationship between the rate of successful attack detection and the magnitude of the sensing attack; ii) human elements, such as the likelihood of human intervention and trust in CAV systems; and iii) heterogeneity in the traffic system, which might involve CAVs operating on different algorithms or models, as well as human-driven vehicles exhibiting varied driving behaviors. By incorporating these factors, the deterministic impact analysis could be expanded into a more realistic, probabilistic framework, which could further inform policy-making and traffic management. In the area of robust CAV control, future research could inspire from this study to consider both communication and sensing attacks in a holistic manner. The focus could be on effectively integrating sensor and communication systems so that if one becomes unreliable, the other can step in to enhance reliability and sustain optimal control performance.

In general, this study provides a comprehensive review of sensing attacks from three perspectives: i) attack models targeting various onboard sensors, ii) potential impact on the target vehicle, and iii) experimental settings to validate the sensing attacks. Based on this review, we have identified a key gap in the current understanding of sensing attacks, which is that most research concentrates on individual target vehicles rather than taking a more macroscopic view, which is of greater concern to transportation agencies.

Therefore, we have discussed extending existing research by considering the systemic implications of sensing attacks through both simulation and field experiments. Building on the characteristics of various types of sensing attacks, we have also designed three representative scenarios to serve as a foundation for future studies, particularly in the areas of impact analysis and robust CAV control. The aim of this study is to serve both as an alert to transportation engineers about the urgent necessity of incorporating sensing attack considerations into their planning and to provide a solid foundation for future studies to understand and mitigate the impact of such attacks at a macroscopic level. Although this study offers an in-depth review of existing sensing attack models and their corresponding impact, it is important to recognize that attack models continue to evolve alongside advancements in sensor technology and algorithms. Therefore, the scope of this overview may not fully capture emerging security threats. Additionally, this study reviews the sensing attacks that influence vehicle behavior, while other passive attacks like eavesdropping and man-in-the-middle are not covered.

ACKNOWLEDGMENT

This research is supported by the Texas A&M Institute of Data Science Career Initiation Fellow Award from Dr. Yang Zhou.

REFERENCES

[1] Y. Zhou, S. Ahn, M. Wang, and S. Hoogendoorn, "Stabilizing mixed vehicular platoons with connected automated vehicles: An H-infinity approach," *Transportation Research Part B: Methodological*, vol. 132, pp. 152–170, Feb. 2020, doi: 10.1016/j.trb.2019.06.005.

[2] Y. Zhou, M. Wang, and S. Ahn, "Distributed model predictive control approach for cooperative car-following with guaranteed local and string stability," *Transportation Research Part B: Methodological*, vol. 128, pp. 69–86, Oct. 2019, doi: 10.1016/j.trb.2019.07.001.

[3] D. Watzenig and M. Horn, "Introduction to automated driving," *Automated driving: Safer and more efficient future driving*, pp. 3–16, 2017.

[4] A. Sarker *et al.*, "A review of sensing and communication, human factors, and controller aspects for information-aware connected and automated vehicles," *IEEE transactions on intelligent transportation systems*, vol. 21, no. 1, pp. 7–29, 2019.

[5] P. Wang, X. Wu, and X. He, "Modeling and analyzing cyberattack effects on connected automated vehicular platoons," *Transportation Research Part C: Emerging Technologies*, vol. 115, p. 102625, Jun. 2020, doi: 10.1016/j.trc.2020.102625.

[6] Y. Li, Y. Tu, Q. Fan, C. Dong, and W. Wang, "Influence of cyber-attacks on longitudinal safety of connected and automated vehicles," *Accident Analysis & Prevention*, vol. 121, pp. 148–156, Dec. 2018, doi: 10.1016/j.aap.2018.09.016.

[7] Z. H. Khattak, B. L. Smith, and M. D. Fontaine, "Impact of cyberattacks on safety and stability of connected and automated vehicle platoons under lane changes," *Accident Analysis & Prevention*, vol. 150, p. 105861, Feb. 2021, doi: 10.1016/j.aap.2020.105861.

[8] Y. Tu, W. Wang, Y. Li, C. Xu, T. Xu, and X. Li, "Longitudinal safety impacts of cooperative adaptive cruise control vehicle's degradation," *Journal of Safety Research*, vol. 69, pp. 177–192, Jun. 2019, doi: 10.1016/j.jsr.2019.03.002.

[9] R. Cheng, H. Lyu, Y. Zheng, and H. Ge, "Modeling and stability analysis of cyberattack effects on heterogeneous intelligent traffic

flow," *Physica A: Statistical Mechanics and its Applications*, vol. 604, p. 127941, Oct. 2022, doi: 10.1016/j.physa.2022.127941.

[10] J. Petit, B. Stottelaar, M. Feiri, and F. Kargl, "Remote attacks on automated vehicles sensors: Experiments on camera and lidar," *Black Hat Europe*, vol. 11, no. 2015, p. 995, 2015.

[11] Y. Cao et al., "Adversarial sensor attack on lidar-based perception in autonomous driving," in *Proceedings of the 2019 ACM SIGSAC conference on computer and communications security*, 2019, pp. 2267–2281.

[12] T. Huang, Y. Chen, B. Yao, B. Yang, X. Wang, and Y. Li, "Adversarial attacks on deep-learning-based radar range profile target recognition," *Information Sciences*, vol. 531, pp. 159–176, 2020.

[13] F. A. Butt, J. N. Chattha, J. Ahmad, M. U. Zia, M. Rizwan, and I. H. Naqvi, "On the integration of enabling wireless technologies and sensor fusion for next-generation connected and autonomous vehicles," *IEEE Access*, vol. 10, pp. 14643–14668, 2022.

[14] Z. El-Rewini, K. Sadatsharan, N. Sugunaraj, D. F. Selvaraj, S. J. Plathottam, and P. Ranganathan, "Cybersecurity attacks in vehicular sensors," *IEEE Sensors Journal*, vol. 20, no. 22, pp. 13752–13767, 2020.

[15] A. Modas, R. Sanchez-Matilla, P. Frossard, and A. Cavallaro, "Toward robust sensing for autonomous vehicles: An adversarial perspective," *IEEE Signal Processing Magazine*, vol. 37, no. 4, pp. 14–23, 2020.

[16] H. Wei et al., "Physical adversarial attack meets computer vision: A decade survey," *arXiv preprint arXiv:2209.15179*, 2022.

[17] T. Sato, S. H. V. Bhupathiraju, M. Clifford, T. Sugawara, Q. A. Chen, and S. Rampazzi, "WIP: Infrared Laser Reflection Attack Against Traffic Sign Recognition Systems," in *Proceedings Inaugural International Symposium on Vehicle Security & Privacy. San Diego, CA, USA: Internet Society*, 2023.

[18] Y. Cao et al., "Invisible for both Camera and LiDAR: Security of Multi-Sensor Fusion based Perception in Autonomous Driving Under Physical-World Attacks," in *2021 IEEE Symposium on Security and Privacy (SP)*, May 2021, pp. 176–194. doi: 10.1109/SP40001.2021.00076.

[19] J. Lu, H. Sibai, E. Fabry, and D. Forsyth, "No need to worry about adversarial examples in object detection in autonomous vehicles," *arXiv preprint arXiv:1707.03501*, 2017.

[20] Y. Guo, C. DiPalma, T. Sato, Y. Cao, Q. A. Chen, and Y. C. NIO, "An Adversarial Attack on DNN-based Adaptive Cruise Control Systems".

[21] H. Shin, D. Kim, Y. Kwon, and Y. Kim, "Illusion and dazzle: Adversarial optical channel exploits against lidars for automotive applications," in *Cryptographic Hardware and Embedded Systems–CHES 2017: 19th International Conference, Taipei, Taiwan, September 25-28, 2017, Proceedings*, Springer, 2017, pp. 445–467.

[22] B. G. Stottelaar, "Practical cyber-attacks on autonomous vehicles," University of Twente, 2015.

[23] R. S. Hallyburton, Y. Liu, Y. Cao, Z. M. Mao, and M. Pajic, "Security Analysis of {Camera-LiDAR} Fusion Against {Black-Box} Attacks on Autonomous Vehicles," in *31st USENIX Security Symposium (USENIX Security 22)*, 2022, pp. 1903–1920.

[24] E. Yeh, J. Choi, N. Prelcic, C. Bhat, and R. W. Heath Jr, "Security in automotive radar and vehicular networks," *submitted to Microwave Journal*, 2016.

[25] R. Poisel, *Modern communications jamming principles and techniques*. Artech house, 2011.

[26] A. Lazaro, A. Porcel, M. Lazaro, R. Villarino, and D. Girbau, "Spoofing attacks on FMCW radars with low-cost backscatter tags," *Sensors*, vol. 22, no. 6, p. 2145, 2022.

[27] R. Komissarov and A. Wool, "Spoofing Attacks Against Vehicular FMCW Radar," in *Proceedings of the 5th Workshop on Attacks and Solutions in Hardware Security*, in ASHES '21. New York, NY, USA: Association for Computing Machinery, Nov. 2021, pp. 91–97. doi: 10.1145/3474376.3487283.

[28] C. Yan, W. Xu, and J. Liu, "Can you trust autonomous vehicles: Contactless attacks against sensors of self-driving vehicle," *Def Con*, vol. 24, no. 8, p. 109, 2016.

[29] W. Xu, C. Yan, W. Jia, X. Ji, and J. Liu, "Analyzing and Enhancing the Security of Ultrasonic Sensors for Autonomous Vehicles," *IEEE Internet of Things Journal*, vol. 5, no. 6, pp. 5015–5029, Dec. 2018, doi: 10.1109/JIOT.2018.2867917.

[30] B. S. Lim, S. L. Keoh, and V. L. L. Thing, "Autonomous vehicle ultrasonic sensor vulnerability and impact assessment," in *2018 IEEE 4th World Forum on Internet of Things (WF-IoT)*, Feb. 2018, pp. 231–236. doi: 10.1109/WF-IoT.2018.8355132.

[31] T. Gluck, M. Kravchik, S. Chocron, Y. Elovici, and A. Shabtai, "Spoofing Attack on Ultrasonic Distance Sensors Using a Continuous Signal," *Sensors*, vol. 20, no. 21, Art. no. 21, Jan. 2020, doi: 10.3390/s20216157.

[32] J. Lou, Q. Yan, Q. Hui, and H. Zeng, "SoundFence: Securing Ultrasonic Sensors in Vehicles Using Physical-Layer Defense," in *2021 18th Annual IEEE International Conference on Sensing, Communication, and Networking (SECON)*, Jul. 2021, pp. 1–9. doi: 10.1109/SECON52354.2021.9491590.

[33] M. Kamal, A. Barua, C. Vitale, C. Laoudias, and G. Ellinas, "GPS Location Spoofing Attack Detection for Enhancing the Security of Autonomous Vehicles," in *2021 IEEE 94th Vehicular Technology Conference (VTC2021-Fall)*, Sep. 2021, pp. 1–7. doi: 10.1109/VTC2021-Fall52928.2021.9625567.

[34] R. A. Agyapong, M. Nabil, A.-R. Nuhu, M. I. Rasul, and A. Homaifar, "Efficient Detection of GPS Spoofing Attacks on Unmanned Aerial Vehicles Using Deep Learning," in *2021 IEEE Symposium Series on Computational Intelligence (SSCI)*, Dec. 2021, pp. 01–08. doi: 10.1109/SSCI50451.2021.9659972.

[35] A. Gasimova, T. T. Khoei, and N. Kaabouch, "A Comparative Analysis of the Ensemble Models for Detecting GPS Spoofing attacks on UAVs," in *2022 IEEE 12th Annual Computing and Communication Workshop and Conference (CCWC)*, Jan. 2022, pp. 0310–0315. doi: 10.1109/CCWC54503.2022.9720738.

[36] C. Vitale et al., "CARAMEL: results on a secure architecture for connected and autonomous vehicles detecting GPS spoofing attacks," *EURASIP Journal on Wireless Communications and Networking*, vol. 2021, no. 1, p. 115, May 2021, doi: 10.1186/s13638-021-01971-x.

[37] C. Sanders and Y. Wang, "Localizing Spoofing Attacks on Vehicular GPS Using Vehicle-to-Vehicle Communications," *IEEE Transactions on Vehicular Technology*, vol. 69, no. 12, pp. 15656–15667, Dec. 2020, doi: 10.1109/TVT.2020.3031576.

[38] Z. Yang et al., "Anomaly Detection Against GPS Spoofing Attacks on Connected and Autonomous Vehicles Using Learning From Demonstration," *IEEE Transactions on Intelligent Transportation Systems*, vol. 24, no. 9, pp. 9462–9475, Sep. 2023, doi: 10.1109/TITS.2023.3269029.

[39] Y. Cao et al., "Adversarial Sensor Attack on LiDAR-based Perception in Autonomous Driving," in *Proceedings of the 2019 ACM SIGSAC Conference on Computer and Communications Security*, in CCS '19. New York, NY, USA: Association for Computing Machinery, Nov. 2019, pp. 2267–2281. doi: 10.1145/3319535.3339815.

[40] P. Kapoor, A. Vora, and K.-D. Kang, "Detecting and Mitigating Spoofing Attack Against an Automotive Radar," in *2018 IEEE 88th Vehicular Technology Conference (VTC-Fall)*, Aug. 2018, pp. 1–6. doi: 10.1109/VTCFall.2018.8690734.

[41] C. You, Z. Hau, and S. Demetriou, "Temporal Consistency Checks to Detect LiDAR Spoofing Attacks on Autonomous Vehicle Perception," in *Proceedings of the 1st Workshop on Security and Privacy for Mobile AI*, in MAISP'21. New York, NY, USA: Association for Computing Machinery, Jul. 2021, pp. 13–18. doi: 10.1145/3469261.3469406.

[42] Z. Sun, S. Balakrishnan, L. Su, A. Bhuyan, P. Wang, and C. Qiao, "Who Is in Control? Practical Physical Layer Attack and Defense for mmWave-Based Sensing in Autonomous Vehicles," *IEEE Transactions on Information Forensics and Security*, vol. 16, pp. 3199–3214, 2021, doi: 10.1109/TIFS.2021.3076287.

[43] Y. Wang, N. Masoud, and A. Khojandi, "Real-time sensor anomaly detection and recovery in connected automated vehicle sensors," *IEEE transactions on intelligent transportation systems*, vol. 22, no. 3, pp. 1411–1421, 2020.

[44] G. Wan et al., "Robust and precise vehicle localization based on multi-sensor fusion in diverse city scenes," in *2018 IEEE international conference on robotics and automation (ICRA)*, IEEE, 2018, pp. 4670–4677.